\documentclass{article}
\pdfoutput=1
\usepackage{lscape}
\usepackage{enumerate}
\usepackage{amsfonts}
\usepackage{amsmath}
\usepackage{textcomp}
\usepackage{amssymb}
\usepackage{color}
\usepackage{graphicx}
\usepackage{pdflscape}
\usepackage{afterpage}
\usepackage[matrix,arrow,curve]{xy}
\usepackage{array}
\usepackage{indentfirst}

\def\be{\begin{equation}}
\def\ee{\end{equation}}

\makeatletter
\renewcommand*{\@cite@ofmt}{\bfseries\hbox}
\makeatother

\def\black{\color{black}}

\hoffset= - 1.2in         

\begin{document}

\title{\vspace{0.1cm}{\Large {\bf  Entangled states from arborescent knots}\vspace{.2cm}}
\author{\bf Sergey Mironov$^{a,c,d}$\thanks{e-mail: sa.mironov\_1@physics.msu.ru},
Andrey Morozov$^{b,c,e}$\thanks{e-mail: morozov.andrey.a@iitp.ru}}
}
\date{ }

\maketitle

\vspace{-5.5cm}

\begin{center}
\hfill ITEP/TH-12/24\\
\hfill IITP/TH-10/24\\
\hfill MIPT/TH-8/24\\
\end{center}

\vspace{3.6cm}

\begin{center}
$^a$ {\small {\it INR RAS, Moscow 117312, Russia}}\\
$^b$ {\small {\it IITP RAS, Moscow 127994, Russia}}\\
$^c$ {\small {\it NRC ``Kurchatov Institute'', Moscow 123182, Russia}}\\
$^d$ {\small {\it ITMP, MSU, Moscow, 119991, Russia}}\\
$^e$ {\small {\it MIPT, Dolgoprudny, 141701, Russia}}\\
\end{center}

\vspace{1cm}

\begin{abstract}
In this paper we discuss how to use arborescent knots to construct entangled multi-qubit states. We show that Bell-states, GHZ-states and cluster states can be constructed from such knots. The latter are particularly interesting since they form a base for the measurement-based quantum computers.
\end{abstract}

\vspace{.5cm}



\section{Introduction}

In \cite{Kit} the model of topological quantum computer was suggested. Qubit states in such a computer are topological and, therefore, are much more stable, than in other models of quantum computers. Operations and quantum states are made by studying quasi-particles, called anyons. These particles are special in a sense that their intertwining produce some non-trivial operators. While experimental realization of non-abelian anyons, required for the topological quantum computer is lacking, there is extensive research in this direction with some promising results \cite{ReviewTQC}-\cite{expend}.

 Anyons are modeled by the Chern-Simons theory with gauge group usually chosen as $SU(N)$:
\begin{equation}
S_{CS}=\frac{k}{4\pi}\int\limits_{S^3}\text{Tr}\left( A\wedge dA+\frac{2}{3}A\wedge A\wedge A\right).
\label{e:SCS}
\end{equation}
Here $k$ is coupling constant which should be integer to preserve topological invariance. 
Anyon trajectories form links and knots in a three-dimensional space. The crossings of these trajectories and, correspondingly, operations in topological quantum computer correspond to the quantum $\mathcal{R}$-matrices \cite{KLJ}.

\begin{figure}[h!]
\begin{picture}(150,140)(-230,-65)
\qbezier(-12,0)(-18,-6)(-18,-12)
\qbezier(-10,-2)(-6,-6)(-6,-12)
\qbezier(10,-2)(6,-6)(6,-12)
\qbezier(12,0)(18,-6)(18,-12)
\put(-18,-12){\line(0,-1){34}}
\qbezier(-6,-12)(-6,-18)(-2,-22)
\qbezier(6,-12)(6,-18)(0,-24)
\put(18,-12){\line(0,-1){34}}
\qbezier(0,-24)(-6,-30)(-6,-36)
\qbezier(2,-26)(6,-30)(6,-36)
\put(-6,-36){\line(0,-1){10}}
\put(6,-36){\line(0,-1){10}}
\qbezier(-18,-46)(-18,-52)(-12,-52)
\qbezier(-6,-46)(-6,-52)(-12,-52)
\qbezier(18,-46)(18,-52)(12,-52)
\qbezier(6,-46)(6,-52)(12,-52)
\qbezier(-14,2)(-18,6)(-18,12)
\qbezier(-12,0)(-6,6)(-6,12)
\qbezier(12,0)(6,6)(6,12)
\qbezier(14,2)(18,6)(18,12)
\put(-18,12){\line(0,1){34}}
\qbezier(-6,12)(-6,18)(-2,22)
\qbezier(6,12)(6,18)(0,24)
\put(18,12){\line(0,1){34}}
\qbezier(0,24)(-6,30)(-6,36)
\qbezier(2,26)(6,30)(6,36)
\put(-6,36){\line(0,1){10}}
\put(6,36){\line(0,1){10}}
\qbezier(-18,46)(-18,52)(-12,52)
\qbezier(-6,46)(-6,52)(-12,52)
\qbezier(18,46)(18,52)(12,52)
\qbezier(6,46)(6,52)(12,52)
\put(-30,41){\line(1,0){60}}
\put(-30,61){\line(0,-1){122}}
\put(30,-41){\line(-1,0){60}}
\put(30,-61){\line(0,1){122}}
\put(-30,61){\line(1,0){60}}
\put(-30,-61){\line(1,0){60}}
\put(-150,50){\hbox{annihilation/measurement}}
\put(-130,0){\hbox{entangling/operations}}
\put(-70,-50){\hbox{creation}}
\put(50,0){\hbox{$\rightarrow$}}
\put(110,0){
\put(-18,-36){\line(0,-1){10}}
\put(18,-36){\line(0,-1){10}}
\put(-6,-36){\line(0,-1){10}}
\put(6,-36){\line(0,-1){10}}
\qbezier(-18,-46)(-18,-52)(-12,-52)
\qbezier(-6,-46)(-6,-52)(-12,-52)
\qbezier(18,-46)(18,-52)(12,-52)
\qbezier(6,-46)(6,-52)(12,-52)
\put(-18,36){\line(0,1){10}}
\put(18,36){\line(0,1){10}}
\put(-6,36){\line(0,1){10}}
\put(6,36){\line(0,1){10}}
\qbezier(-18,46)(-18,52)(-12,52)
\qbezier(-6,46)(-6,52)(-12,52)
\qbezier(18,46)(18,52)(12,52)
\qbezier(6,46)(6,52)(12,52)
\put(-24,-36){\line(1,0){48}}
\put(-24,-36){\line(0,1){72}}
\put(24,36){\line(-1,0){48}}
\put(24,36){\line(0,-1){72}}
\put(-5,-5){\hbox{$\mathcal{B}$}}
}
\end{picture}
\caption{Description of one-qubit operations using anyons. Two pairs of anyons are created then they are entangled and then annihilated. Trajectories of anyons form a knot or a link. The braid/entangling of two pairs of anyons corresponds to an operator $\mathcal{B}$ which is a two by two matrix.\label{f:anyons}}
\end{figure}
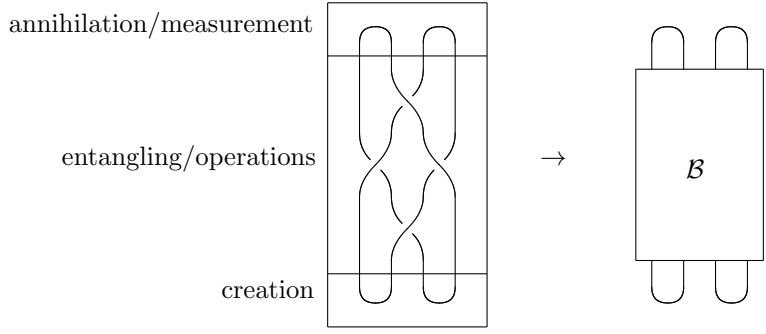

As was extensively discussed in \cite{TopoBook,TowTQC,QuantKnots,LargeK,Measure}, one qubit corresponds to the state of two anyon-anti-anyon pairs, like on Fig. \ref{f:anyons} (see \cite{TopoBook} for details). We will further discuss where the corresponding Hilbert space appears from and how the quantum operations look like in Section \ref{s:1qb}.

In \cite{QuantKnots,LargeK} we've shown that quantum $\mathcal{R}$-matrices form a universal set of one-qubit gates for large enough values of $k$. And operations on the anyons can also be used to measure $k$ itself \cite{Bel}. However for effective universal quantum computer we also need two-qubit gates, and they are, as usual, much harder to construct. There is no known unitary realization of universal two-qubit gates in topological quantum computer \cite{TopoBook}, since all of them require some kind of measurement or can take the system out of computational space and thus cause leakage errors for quantum computations. In practice this leakage can be heavily reduced by additional sequence of braidings \cite{TopoBook}, however, as far as we understand there is no approach, which allows construction of universal set of gates with zero leakage. But we want to discuss different approach to the topological quantum computer, which is closer to the measurement-based quantum computers.

In this paper we discuss how arborescent knots can be interpreted as two-qubit operations and how we can construct Bell-states and cluster states using this approach. These two-qubit operations are still non-unitary and thus not allowing a quantum circuit interpretation of the topological quantum computer. In other words, leakage problem is still present and quantum mechanical probability in computational space is not conserved. However, cluster states \cite{clusterstate}, which can be constructed using these two-qubit operations provide a base for construction of a measurement-based quantum computer \cite{QCc}. Such a computer is constructed by applying one-qubit quantum measurements to a complex cluster state. In this paper we show, that arborescent knot operations indeed provide one with possibility to construct the desired cluster state. In particular, to have many-qubit operations one needs a two dimensional cluster state (with non linear graph) \cite{Measurement-based} and such cluster states can also be provided by arborescent knots. While cluster states are of course a known structure, in this paper we discuss how to get them on the topological quantum computer, which can lead to the construction of topological measurement based quantum computer in the future.

\section{One-qubit operations \label{s:1qb}}

Two-bridge knots, like on Fig. \ref{f:anyons}, in fundamental representation can be calculated using a simple set of operators. Quantum $\mathcal{R}$-matrices do not change the irreducible representations travelling along any number of strands. When pair of anyons is created, they appear in trivial representation. This means that the whole braid is in trivial representation.

However, there are multiple trivial representations appearing in such a braid - two-dimensional space of those. We can get them for example as follows. In the basis at the bottom of the braid on Fig. \ref{f:anyons}, the first of these trivial representations appear when the pair of anyons on the left is in the trivial representation (the pair as a whole, not each of them), therefore the pair on the right is also in a trivial representation. The second trivial representation appears when both pairs are in adjoint representation.
\begin{equation}
\begin{array}{l}
|V_{\emptyset}>:\ \ \ ([1]\otimes\bar{[1]})\otimes([1]\otimes\bar{[1]})\rightarrow\emptyset\otimes\emptyset\rightarrow\emptyset,
\\
|V_1>:\ \ \ ([1]\otimes\bar{[1]})\otimes([1]\otimes\bar{[1]})\rightarrow\text{adj}\otimes\text{adj}\rightarrow\emptyset.
\end{array}
\end{equation}
Vectors $|V_{\emptyset}>$ and $|V_1>$ form a Hilbert space of one qubit. Let us mention again, that this two-dimensional space of a qubit (or d-dimensional in case of qudit) is not an internal space of some irreducible representation, but the space of different irreducible representations itself. The closest analogy are $SU(3)$ and isospin multiplets of particles. Anyons in different representations correspond to particles in different hypermultiplets, in other words, different particles.
\black

There are four two-dimensional operators, acting on a qubit \cite{QuantKnots,Tabul,DoubleFat,NPZ,MMS}:
 \begin{equation}
\begin{array}{lcrclcr}
S&=&
\frac{1}{\sqrt{(q+q^{-1})(A-A^{-1})}}\left(\begin{array}{cc}\sqrt{\frac{A}{q}-\frac{q}{A}} & \sqrt{Aq-\frac{1}{Aq}} \\ \sqrt{Aq-\frac{1}{Aq}} & -\sqrt{\frac{A}{q}-\frac{q}{A}} \end{array}\right);
& &
T&=&\left(\begin{array}{cc}
\frac{q}{A} \\ & -\frac{1}{qA}
\end{array}\right);
\\ \\
\bar{S}&=&
\left(\begin{array}{cc}\frac{q-q^{-1}}{A-A^{-1}} & \frac{\sqrt{(Aq-\frac{1}{Aq})(\frac{A}{q}-\frac{q}{A})}}{A-A^{-1}} \\ \frac{\sqrt{(Aq-\frac{1}{Aq})(\frac{A}{q}-\frac{q}{A})}}{A-A^{-1}} & -\frac{q-q^{-1}}{A-A^{-1}} \end{array}\right);
 & &
\bar{T}&=&\left(\begin{array}{cc}
1 \\ & -A
\end{array}\right),
\end{array}
\label{eq:ST}
\end{equation}
where $q=\text{exp}(\frac{2\pi i}{k+N})$ and $A=q^N$ with $k$ and $N$ being parameters of Chern-Simons theory (\ref{e:SCS}).
These operators correspond to the braiding of different strands ($T$-operators) or change of bases ($S$-operators).
As was discussed in \cite{LargeK} these operators form a universal set of one-qubit gates, and their generalization to higher representations can give us even universal one-qudit gates.

From the braid the product of operators corresponding to the crossings in the braid is constructed, see \cite{LargeK} for details. Thus for each braid we can write a two-by-two matrix $\mathcal{B}_{XY}$, which is a product of matrices from (\ref{eq:ST}) in an order defined by a braid. Then the process like on Fig. \ref{f:anyons} corresponds to the matrix element of $\mathcal{B}$ with $X$ and $Y$ equal to the trivial representations. From the point of view of anyons, $\mathcal{B}_{\emptyset\emptyset}$ describes the probability amplitude of two pairs of anyons being created, then entangled and then annihilated. If they do not annihilate then this corresponds to the element $\mathcal{B}_{\emptyset\text{adj}}$.

Therefore, creation of initial state of quantum computer corresponds to the birth of two anyon-anti-anyon pairs. Then this state can be changed by applying operators from (\ref{eq:ST}). Finally we measure the state by trying to annihilate these pairs.

Since we have universal set of one-qubit gates, further in the text we will presume that matrix $\mathcal{B}$ can be any unitary matrix.

\section{Arborescent knots}

\begin{figure}[h!]
\begin{picture}(200,160)(-125,-75)
%
\put(0,0){
\put(0,0){\line(0,1){16}}
\put(0,0){\line(1,0){25}}
\put(25,16){\line(0,-1){16}}
\put(25,16){\line(-1,0){25}}
\put(33,-33){\line(1,0){16}}
\put(33,-33){\line(0,1){25}}
\put(49,-8){\line(-1,0){16}}
\put(49,-8){\line(0,-1){25}}
\put(33,24){\line(1,0){16}}
\put(33,24){\line(0,1){25}}
\put(49,49){\line(-1,0){16}}
\put(49,49){\line(0,-1){25}}
\put(57,0){\line(0,1){16}}
\put(57,0){\line(1,0){25}}
\put(82,16){\line(0,-1){16}}
\put(82,16){\line(-1,0){25}}
\put(-13,-6){\line(-1,1){11}}
\put(-13,-6){\line(-1,-1){16}}
\put(-40,-11){\line(1,-1){11}}
\put(-40,-11){\line(1,1){16}}
\put(-13,22){\line(-1,-1){11}}
\put(-13,22){\line(-1,1){16}}
\put(-40,27){\line(1,1){11}}
\put(-40,27){\line(1,-1){16}}
\qbezier(82,2)(86,2)(86,4)
\qbezier(82,6)(86,6)(86,4)
\qbezier(82,10)(86,10)(86,12)
\qbezier(82,14)(86,14)(86,12)
\qbezier(35,-33)(35,-37)(37,-37)
\qbezier(39,-33)(39,-37)(37,-37)
\qbezier(43,-33)(43,-37)(45,-37)
\qbezier(47,-33)(47,-37)(45,-37)
\qbezier(35,49)(35,53)(37,53)
\qbezier(39,49)(39,53)(37,53)
\qbezier(43,49)(43,53)(45,53)
\qbezier(47,49)(47,53)(45,53)
\qbezier(0,2)(-9,2)(-15,-4)
\qbezier(0,6)(-9,6)(-17,-2)
\qbezier(0,10)(-9,10)(-17,18)
\qbezier(0,14)(-9,14)(-15,20)
\qbezier(-20,1)(-11,8)(-20,15)
\qbezier(-22,3)(-15,8)(-22,13)
\qbezier(-38,29)(-41,32)(-40,33)
\qbezier(-36,31)(-39,34)(-40,33)
\qbezier(-33,34)(-36,37)(-35,38)
\qbezier(-31,36)(-34,39)(-35,38)
\qbezier(-38,-13)(-41,-16)(-40,-17)
\qbezier(-36,-15)(-39,-18)(-40,-17)
\qbezier(-33,-18)(-36,-21)(-35,-22)
\qbezier(-31,-20)(-34,-23)(-35,-22)
\qbezier(25,2)(35,2)(35,-8)
\qbezier(25,6)(39,6)(39,-8)
\qbezier(25,10)(39,10)(39,24)
\qbezier(25,14)(35,14)(35,24)
\qbezier(57,2)(47,2)(47,-8)
\qbezier(57,6)(43,6)(43,-8)
\qbezier(57,14)(47,14)(47,24)
\qbezier(57,10)(43,10)(43,24)
\put(10,-60){\hbox{a).}}
}
\put(250,0){
\put(0,0){\line(0,1){16}}
\put(0,0){\line(1,0){25}}
\put(25,16){\line(0,-1){16}}
\put(25,16){\line(-1,0){25}}
\put(57,0){\line(0,1){16}}
\put(57,0){\line(1,0){25}}
\put(82,16){\line(0,-1){16}}
\put(82,16){\line(-1,0){25}}
\qbezier(82,2)(86,2)(86,4)
\qbezier(82,6)(86,6)(86,4)
\qbezier(82,10)(86,10)(86,12)
\qbezier(82,14)(86,14)(86,12)
\put(-57,0){\line(0,1){16}}
\put(-57,0){\line(1,0){25}}
\put(-32,16){\line(0,-1){16}}
\put(-32,16){\line(-1,0){25}}
\qbezier(-57,2)(-61,2)(-61,4)
\qbezier(-57,6)(-61,6)(-61,4)
\qbezier(-57,10)(-61,10)(-61,12)
\qbezier(-57,14)(-61,14)(-61,12)
\put(33,24){\line(1,0){16}}
\put(33,24){\line(0,1){25}}
\put(49,49){\line(-1,0){16}}
\put(49,49){\line(0,-1){25}}
\put(-24,24){\line(1,0){16}}
\put(-24,24){\line(0,1){25}}
\put(-8,49){\line(-1,0){16}}
\put(-8,49){\line(0,-1){25}}
\qbezier(25,10)(39,10)(39,24)
\qbezier(25,14)(35,14)(35,24)
\qbezier(57,14)(47,14)(47,24)
\qbezier(57,10)(43,10)(43,24)
\qbezier(-32,10)(-18,10)(-18,24)
\qbezier(-32,14)(-22,14)(-22,24)
\qbezier(0,14)(-10,14)(-10,24)
\qbezier(0,10)(-14,10)(-14,24)
\put(25,2){\line(1,0){32}}
\put(25,6){\line(1,0){32}}
\put(-32,2){\line(1,0){32}}
\put(-32,6){\line(1,0){32}}
\qbezier(-10,49)(-10,60)(12.5,60)
\qbezier(35,49)(35,60)(12.5,60)
\qbezier(-14,49)(-14,64)(12.5,64)
\qbezier(39,49)(39,64)(12.5,64)
\qbezier(-18,49)(-18,68)(12.5,68)
\qbezier(43,49)(43,68)(12.5,68)
\qbezier(-22,49)(-22,72)(12.5,72)
\qbezier(47,49)(47,72)(12.5,72)
{\color{red}
\qbezier(25,8)(41,8)(41,24)
\qbezier(0,8)(-16,8)(-16,24)
\qbezier(12.5,66)(-16,66)(-16,49)
\qbezier(12.5,66)(41,66)(41,49)
\put(0,8){\line(1,0){25}}
\put(41,24){\line(0,1){25}}
\put(-16,24){\line(0,1){25}}
}
\put(10,-60){\hbox{b).}}
}
\end{picture}
\caption{On figure a). an example of arborescent knot is shown. It is a tree made from four-strand blocks. On figure b). an example of non-arborescent knot is presented. It includes a loop, marked by the red line.\label{f:tree}}
\end{figure}
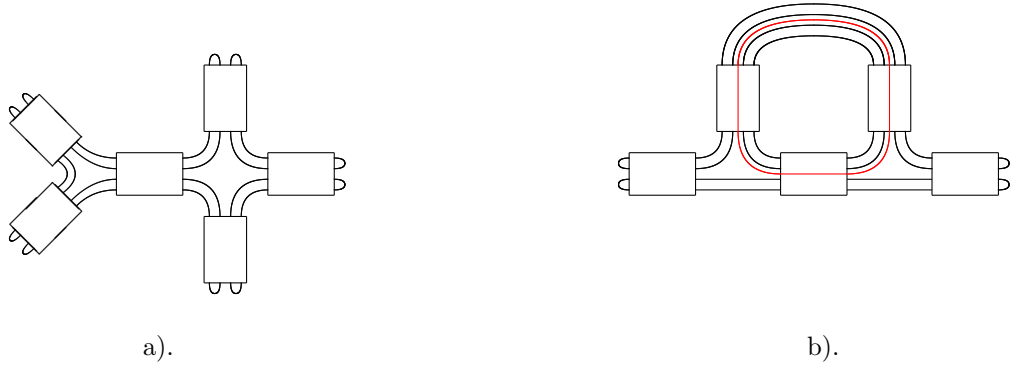

Unclosed braids can form the constructing blocks for a more evolved type of knots called arborescent knots. These blocks has two pairs of strands on each side with which it can be connected to other blocks. From these blocks a tree can be constructed, see Fig. \ref{f:tree}a). In vertices of the tree several such blocks are connected and at the end of the ``branches'' braids are closed. If there are no loops, like on Fig. \ref{f:tree}b)., each braid still carries trivial representation. This means that operators from (\ref{eq:ST}) are enough to construct the polynomials of such knots.

Knot polynomials are constructed from matrices $\mathcal{B}$ corresponding to the building blocks using the following procedure.  Each matrix $\mathcal{B}$ has two indices corresponding to the ends of the braid.
All matrices $\mathcal{B}$ should be multiplied, but their indices should be dealt with accordingly to the picture.
If braid is closed, the corresponding index should be zero.
Vertices, where several braids are connected, correspond to a summation over the indices in this vertex, \cite{Tabul,DoubleFat}:
\begin{equation}
\mathcal{V}:\ \ \sum\limits_X (d_{X})^n\prod\limits_{i=1}^n \mathcal{B}^{(i)}_{X Y_i},
\end{equation}
where $n$ is a number of braids, connecting in the vertex, $i$ enumerate these braids, $d_{X}$ -- is a dimension of representation $X$. $d_{X}=(S_{\emptyset X})^{-1}$ or $d_{X}=(\bar{S}_{\emptyset X})^{-1}$, depending on the direction of the strands.

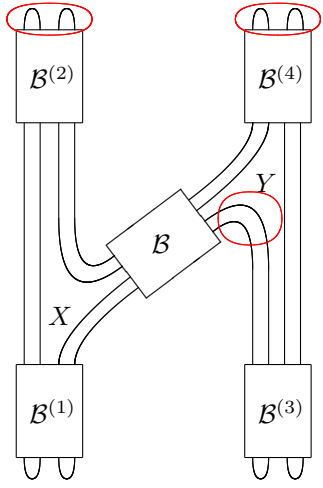
\begin{figure}[h!]
\begin{picture}(200,200)(-175,-75)
\put(0,-50){
\put(0,0){\line(1,0){25}}
\put(0,0){\line(0,1){35}}
\put(25,35){\line(-1,0){25}}
\put(25,35){\line(0,-1){35}}
\put(5,13){\hbox{$\mathcal{B}^{(1)}$}}
\qbezier(3,0)(3,-8)(6,-8)
\qbezier(9,0)(9,-8)(6,-8)
\qbezier(16,0)(16,-8)(19,-8)
\qbezier(22,0)(22,-8)(19,-8)
}
\put(0,76){
\put(0,0){\line(1,0){25}}
\put(0,0){\line(0,1){35}}
\put(25,35){\line(-1,0){25}}
\put(25,35){\line(0,-1){35}}
\put(5,13){\hbox{$\mathcal{B}^{(2)}$}}
\qbezier(3,35)(3,43)(6,43)
\qbezier(9,35)(9,43)(6,43)
\qbezier(16,35)(16,43)(19,43)
\qbezier(22,35)(22,43)(19,43)
}
\put(34,10){
\put(15,0){\line(-3,4){15}}
\put(15,0){\line(4,3){28}}
\put(28,41){\line(3,-4){15}}
\put(28,41){\line(-4,-3){28}}
\put(17,17){\hbox{$\mathcal{B}$}}
}
\put(86,-50){
\put(0,0){\line(1,0){25}}
\put(0,0){\line(0,1){35}}
\put(25,35){\line(-1,0){25}}
\put(25,35){\line(0,-1){35}}
\put(5,13){\hbox{$\mathcal{B}^{(3)}$}}
\qbezier(3,0)(3,-8)(6,-8)
\qbezier(9,0)(9,-8)(6,-8)
\qbezier(16,0)(16,-8)(19,-8)
\qbezier(22,0)(22,-8)(19,-8)
}
\put(86,76){
\put(0,0){\line(1,0){25}}
\put(0,0){\line(0,1){35}}
\put(25,35){\line(-1,0){25}}
\put(25,35){\line(0,-1){35}}
\put(5,13){\hbox{$\mathcal{B}^{(4)}$}}
\qbezier(3,35)(3,43)(6,43)
\qbezier(9,35)(9,43)(6,43)
\qbezier(16,35)(16,43)(19,43)
\qbezier(22,35)(22,43)(19,43)
}
\qbezier(16,40)(16,4)(40,22)
\qbezier(22,40)(22,14)(37,26)
\qbezier(16,-15)(16,-3)(43,18)
\qbezier(22,-15)(22,-4)(46,14)
\put(111,61){
\qbezier(-16,-40)(-16,-4)(-40,-22)
\qbezier(-22,-40)(-22,-14)(-37,-26)
\qbezier(-16,15)(-16,3)(-43,-18)
\qbezier(-22,15)(-22,4)(-46,-14)
}
\put(9,-15){\line(0,1){91}}
\put(3,-15){\line(0,1){91}}
\put(107,-15){\line(0,1){91}}
\put(101,-15){\line(0,1){91}}
\put(16,40){\line(0,1){36}}
\put(22,40){\line(0,1){36}}
\put(89,-15){\line(0,1){36}}
\put(95,-15){\line(0,1){36}}
\put(12,0){\hbox{$X$}}
\put(90,50){\hbox{$Y$}}
{\color{red}
\put(12.5,115){
\qbezier(0,-6)(16,-6)(16,0)
\qbezier(0,6)(16,6)(16,0)
\qbezier(0,-6)(-16,-6)(-16,0)
\qbezier(0,6)(-16,6)(-16,0)
}
\put(98.5,115){
\qbezier(0,-6)(16,-6)(16,0)
\qbezier(0,6)(16,6)(16,0)
\qbezier(0,-6)(-16,-6)(-16,0)
\qbezier(0,6)(-16,6)(-16,0)
}
\put(88,40){
\qbezier(0,-10)(12,-10)(12,0)
\qbezier(0,10)(12,10)(12,0)
\qbezier(0,-10)(-12,-10)(-12,0)
\qbezier(0,10)(-12,10)(-12,0)
}
}
\end{picture}
\caption{Simple example of arborescent knots with two vertices. One vertex corresponds to a summation over $X$ and the other -- over $Y$. Left and right column correspond to two qubits. Bridge describes a two-qubit operation. Red curves encircle points where anyons should annihilate.
\label{f:stree}}
\end{figure}

Let us look at the example of a simple tree with two vertices, like on Fig. \ref{f:stree}. The knot polynomial for such arborescent knot is equal to
\begin{equation}
H=\sum\limits_{X,Y} \frac{1}{d_{X}d_{Y}}\mathcal{B}^{(1)}_{\emptyset X}\mathcal{B}^{(2)}_{\emptyset X}\mathcal{B}_{XY}\mathcal{B}^{(3)}_{\emptyset Y}\mathcal{B}^{(4)}_{\emptyset Y}.
\label{eq:Harb}
\end{equation}

Let us also discuss what this polynomial means from the point of view of anyons. When we connect two four strand braids on the left and on the right with a ``bridge'' $\mathcal{B}$ on Fig. \ref{f:stree}, we add two additional points where anyon pairs are created and annihilate. This means that $H$ from (\ref{eq:Harb}) is a probability amplitude of six pairs of anyons created and then all of them annihilating after entangling.

\section{Two-qubit operation from arborescent knots}

If we consider 4-strand braids on the left and on the right of Fig. \ref{f:stree} as two qubits, then the bridge between them can be interpreted as a two qubit operation. Operators $\mathcal{B}^{(1)}$ and $\mathcal{B}^{(3)}$ create initial states for left and right qubits correspondingly, while $\mathcal{B}^{(2)}$ and $\mathcal{B}^{(4)}$ change them to the measurement bases at the end. Therefore, after application of the operators $\mathcal{B}^{(1)}$ and $\mathcal{B}^{(3)}$ we get some two qubit state $|v_{12}>=|v_1>\otimes|v_2>$, where
\begin{equation}
|v_1>=\mathcal{B}^{(1)}_{\emptyset \emptyset}|V_{\emptyset}>+\mathcal{B}^{(1)}_{\emptyset 1}|V_{1}>,\ \ \ |v_2>=\mathcal{B}^{(3)}_{\emptyset \emptyset}|V_{\emptyset}>+\mathcal{B}^{(3)}_{\emptyset 1}|V_{1}>.
\end{equation}
For simplicity of further explanations let us introduce following notation
\begin{equation}
|v_{12}>=a_{\emptyset\emptyset}|V_{\emptyset}>\otimes|V_{\emptyset}>+a_{\emptyset 1}|V_{\emptyset}>\otimes|V_1>+a_{1\emptyset}|V_1>\otimes|V_{\emptyset}>+a_{11}|V_1>\otimes|V_1>.
\end{equation}
From (\ref{eq:Harb}) it can be seen that the resulting state after the bridge will be equal to
\begin{equation}
\begin{array}{l}
|\tilde{v}_{12}>\equiv\mathcal{O}|{v}_{12}>=\cfrac{\mathcal{B}_{XY}}{d_X d_Y}|v_{12}>=
\\ \\
=\cfrac{\mathcal{B}_{\emptyset\emptyset}}{d_{\emptyset}d_{\emptyset}}a_{\emptyset\emptyset}|V_{\emptyset}>\otimes|V_{\emptyset}>+
\cfrac{\mathcal{B}_{\emptyset 1}}{d_{\emptyset}d_{1}}a_{\emptyset 1}|V_{\emptyset}>\otimes|V_1>+
\cfrac{\mathcal{B}_{1\emptyset}}{d_{\emptyset}d_{1}}a_{1\emptyset}|V_1>\otimes|V_{\emptyset}>+
\cfrac{\mathcal{B}_{11}}{d_{1}d_{1}}a_{11}|V_1>\otimes|V_1>.
\end{array}
\end{equation}
\black
This operation corresponds to the following four by four matrix
\begin{equation}
\mathcal{O}=\left(
\begin{array}{cccc}
\frac{\mathcal{B}_{00}}{d_{0}d_{0}} \\
& \frac{\mathcal{B}_{01}}{d_{0}d_{1}} \\
& & \frac{\mathcal{B}_{10}}{d_{0}d_{1}} \\
& & & \frac{\mathcal{B}_{11}}{d_{1}d_{1}} \\
\end{array}\right).
\end{equation}
This matrix is non-unitary, in fact it's determinant is less than $1$, because this operation involves measurement -- projection on the trivial representation when pairs of anyons annihilate at the upper end of the bridge. From the point of view of anyons, this means that when trying to model such operations, we should keep only the instances when the corresponding pairs of anyons annihilate and discard all other instances. In practice it means that at a certain time ($Y$ on the Fig.\ref{f:stree}) one tries to annihilate anyons after braiding; experiment proceeds if anyons annihilate as the picture requires. If they do not annihilate - this could happen is they are in different representations - the experiment stops and one starts over. After such operation, since we discarded all the instances which do not create operator $\mathcal{O}$, we should renormalize the resulting state so that it's norm is again equal to one. If we want operator $\mathcal{O}$ to be linear, this requires norms of all its elements to be equal to each other, then the operator can be easily renormalized. This linear operator is still not unitary and hence can not be used in quantum scheme as a gate. Moreover, we do not have any other two-qubit operation in the construction. However, it turns out that this $\mathcal{O}$ is good enough to construct initial states for measurement based quantum computer.

\section{Creation of Bell states}
One of the easily renormalizable operators $\mathcal{O}$ can be used to create Bell states. Bell states are the most entangled two-qubit states, defined as
\begin{equation}
\begin{array}{lcl}
|\Phi^+\rangle &=& \frac{1}{\sqrt{2}} (|00\rangle + |11\rangle),
\\
|\Phi^-\rangle &=& \frac{1}{\sqrt{2}} (|00\rangle - |11\rangle),
\\
|\Psi^+\rangle &=& \frac{1}{\sqrt{2}} (|01\rangle + |10\rangle),
\\
|\Psi^-\rangle &=& \frac{1}{\sqrt{2}} (|01\rangle - |10\rangle).
\end{array}
\end{equation}

If we take following operations in (\ref{eq:Harb}):
\begin{equation}
\mathcal{B}^{(1)}=\mathcal{B}^{(3)}=\mathcal{B}^{(i)}=\frac{1}{\sqrt{2}}\left(\begin{array}{cc} 1 & 1 \\ -1 & 1\end{array}\right),
\ \ \ \mathcal{B}=\left(\begin{array}{cc} 0 & 1 \\ 1 & 0\end{array}\right),
\end{equation}
then renormalized operator becomes
\begin{equation}
\tilde{\mathcal{O}}=\left(
\begin{array}{cccc}
0 \\
& 1 \\
& & 1 \\
& & & 0 \\
\end{array}\right).
\end{equation}
This means that after applying one-qubit operations $\mathcal{B}^{(1)}$ and $\mathcal{B}^{(3)}$ and two-qubit operation  $\tilde{\mathcal{O}}$ we get Bell state $|\Psi^{+}\rangle$. All other Bell states can be easily constructed from this one using further one-qubit operations. If we continue to apply such two-qubit operations between these qubits and other additional qubits we can get Greenberger–Horne–Zeilinger (GHZ) state. For example, for three qubits we get
\begin{equation}
|\Psi_{123}\rangle=\frac{1}{\sqrt{2}}\left(|010\rangle+|101\rangle\right).
\end{equation}

\section{Cluster states}

Besides Bell and GHZ also other entangled multi-qubit states can be constructed using slight modifications of the approach discussed in the previous section. Among those most interesting are cluster states, which are used in the measurement based computing \cite{clusterstate,QCc,Measurement-based}.

\begin{figure}[h!]
\begin{picture}(200,350)(-175,-75)
\put(0,-50){
\put(0,0){\line(1,0){25}}
\put(0,0){\line(0,1){35}}
\put(25,35){\line(-1,0){25}}
\put(25,35){\line(0,-1){35}}
\put(5,13){\hbox{$\mathcal{B}^{(i)}$}}
\qbezier(3,0)(3,-8)(6,-8)
\qbezier(9,0)(9,-8)(6,-8)
\qbezier(16,0)(16,-8)(19,-8)
\qbezier(22,0)(22,-8)(19,-8)
}

\put(34,10){
\put(15,0){\line(-3,4){15}}
\put(15,0){\line(4,3){28}}
\put(28,41){\line(3,-4){15}}
\put(28,41){\line(-4,-3){28}}
\put(17,17){\hbox{$\mathcal{B}$}}
}
\put(86,-50){
\put(0,0){\line(1,0){25}}
\put(0,0){\line(0,1){35}}
\put(25,35){\line(-1,0){25}}
\put(25,35){\line(0,-1){35}}
\put(5,13){\hbox{$\mathcal{B}^{(i)}$}}
\qbezier(3,0)(3,-8)(6,-8)
\qbezier(9,0)(9,-8)(6,-8)
\qbezier(16,0)(16,-8)(19,-8)
\qbezier(22,0)(22,-8)(19,-8)
}

\qbezier(16,40)(16,4)(40,22)
\qbezier(22,40)(22,14)(37,26)
\qbezier(16,-15)(16,-3)(43,18)
\qbezier(22,-15)(22,-4)(46,14)
\put(111,61){
\qbezier(-16,-40)(-16,-4)(-40,-22)
\qbezier(-22,-40)(-22,-14)(-37,-26)
\qbezier(-16,15)(-16,3)(-43,-18)
\qbezier(-22,15)(-22,4)(-46,-14)
}

\put(16,40){\line(0,1){36}}
\put(22,40){\line(0,1){36}}
\put(89,-15){\line(0,1){36}}
\put(95,-15){\line(0,1){36}}
\put(9,-15){\line(0,1){126}}
\put(3,-15){\line(0,1){126}}
\put(107,-15){\line(0,1){91}}
\put(101,-15){\line(0,1){91}}

\put(86,76){
\put(0,0){\line(1,0){25}}
\put(0,0){\line(0,1){35}}
\put(25,35){\line(-1,0){25}}
\put(25,35){\line(0,-1){35}}
\put(5,13){\hbox{$\mathcal{B}^{(i)}$}}

}
\put(120,136){
\put(15,0){\line(-3,4){15}}
\put(15,0){\line(4,3){28}}
\put(28,41){\line(3,-4){15}}
\put(28,41){\line(-4,-3){28}}
\put(17,17){\hbox{$\mathcal{B}$}}
}
\put(172,76){
\put(0,0){\line(1,0){25}}
\put(0,0){\line(0,1){35}}
\put(25,35){\line(-1,0){25}}
\put(25,35){\line(0,-1){35}}
\put(5,13){\hbox{$\mathcal{B}^{(i)}$}}
\qbezier(3,0)(3,-8)(6,-8)
\qbezier(9,0)(9,-8)(6,-8)
\qbezier(16,0)(16,-8)(19,-8)
\qbezier(22,0)(22,-8)(19,-8)
}
\qbezier(102,166)(102,130)(126,148)
\qbezier(108,166)(108,140)(123,152)
\qbezier(102,111)(102,123)(129,144)
\qbezier(108,111)(108,122)(132,140)
\put(197,187){
\qbezier(-16,-40)(-16,-4)(-40,-22)
\qbezier(-22,-40)(-22,-14)(-37,-26)
\qbezier(-16,15)(-16,3)(-43,-18)
\qbezier(-22,15)(-22,4)(-46,-14)
}

\put(102,166){\line(0,1){106}}
\put(108,166){\line(0,1){106}}
\put(175,111){\line(0,1){36}}
\put(181,111){\line(0,1){36}}
\put(9,111){\line(0,1){161}}
\put(3,111){\line(0,1){161}}
\put(16,76){\line(0,1){196}}
\put(22,76){\line(0,1){196}}
\put(95,111){\line(0,1){161}}
\put(89,111){\line(0,1){161}}
\put(187,111){\line(0,1){91}}
\put(193,111){\line(0,1){91}}
\put(187,237){\line(0,1){36}}
\put(193,237){\line(0,1){36}}
\put(175,237){\line(0,1){36}}
\put(181,237){\line(0,1){36}}
%
\put(172,202){
\put(0,0){\line(1,0){25}}
\put(0,0){\line(0,1){35}}
\put(25,35){\line(-1,0){25}}
\put(25,35){\line(0,-1){35}}
\put(5,13){\hbox{$\mathcal{B}^{(i)}$}}

}
{\color{blue}
\put(-20,0){
\put(0,0){\line(1,0){150}}
\put(0,0){\line(0,1){120}}
\put(150,120){\line(-1,0){150}}
\put(150,120){\line(0,-1){120}}
\put(5,13){\hbox{$C$}}
}
}
{\color{blue}
\put(60,120){
\put(0,0){\line(1,0){150}}
\put(0,0){\line(0,1){120}}
\put(150,120){\line(-1,0){150}}
\put(150,120){\line(0,-1){120}}
\put(5,13){\hbox{$C$}}
}
}
\end{picture}
\caption{Arborescent tangle, used for the construction of linear cluster states. By repeating addition of block $C$ one can get linear cluster state of any size. If block $C$ is instead applied to the qubit in the middle, rather than the most right one, the we get two-dimensional cluster state.
\label{f:manyb}}
\end{figure}
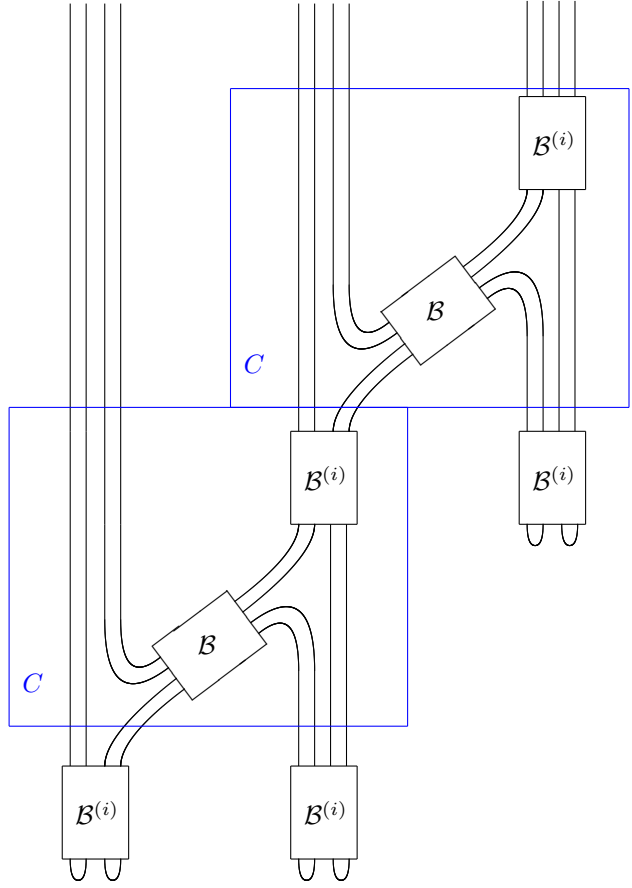

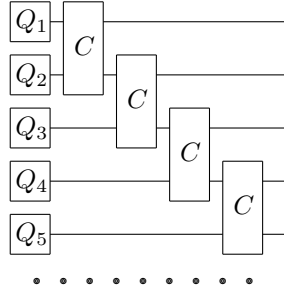
\begin{figure}[h!]
\begin{picture}(200,150)(-175,0)

\put(0,0){
\put(0,0){\line(1,0){15}}
\put(0,0){\line(0,1){15}}
\put(15,15){\line(-1,0){15}}
\put(15,15){\line(0,-1){15}}
\put(2,5){\hbox{$Q_{5}$}}
}
\put(0,20){
\put(0,0){\line(1,0){15}}
\put(0,0){\line(0,1){15}}
\put(15,15){\line(-1,0){15}}
\put(15,15){\line(0,-1){15}}
\put(2,5){\hbox{$Q_{4}$}}
}

\put(0,40){
\put(0,0){\line(1,0){15}}
\put(0,0){\line(0,1){15}}
\put(15,15){\line(-1,0){15}}
\put(15,15){\line(0,-1){15}}
\put(2,5){\hbox{$Q_{3}$}}
}
\put(0,60){
\put(0,0){\line(1,0){15}}
\put(0,0){\line(0,1){15}}
\put(15,15){\line(-1,0){15}}
\put(15,15){\line(0,-1){15}}
\put(2,5){\hbox{$Q_{2}$}}
}
\put(0,80){
\put(0,0){\line(1,0){15}}
\put(0,0){\line(0,1){15}}
\put(15,15){\line(-1,0){15}}
\put(15,15){\line(0,-1){15}}
\put(2,5){\hbox{$Q_{1}$}}
}
\put(15,67.5){\line(1,0){5}}
\put(15,87.5){\line(1,0){5}}
\put(20,60){
\put(0,0){\line(1,0){15}}
\put(0,0){\line(0,1){35}}
\put(15,35){\line(-1,0){15}}
\put(15,35){\line(0,-1){35}}
\put(4,15){\hbox{$C$}}
}
\put(35,87.5){\line(1,0){70}}
\put(35,67.5){\line(1,0){5}}
\put(15,47.5){\line(1,0){25}}
\put(40,40){
\put(0,0){\line(1,0){15}}
\put(0,0){\line(0,1){35}}
\put(15,35){\line(-1,0){15}}
\put(15,35){\line(0,-1){35}}
\put(4,15){\hbox{$C$}}
}
\put(55,67.5){\line(1,0){50}}
\put(55,47.5){\line(1,0){5}}
\put(15,27.5){\line(1,0){45}}
\put(60,20){
\put(0,0){\line(1,0){15}}
\put(0,0){\line(0,1){35}}
\put(15,35){\line(-1,0){15}}
\put(15,35){\line(0,-1){35}}
\put(4,15){\hbox{$C$}}
}
\put(75,47.5){\line(1,0){30}}
\put(75,27.5){\line(1,0){5}}
\put(15,7.5){\line(1,0){65}}
\put(80,0){
\put(0,0){\line(1,0){15}}
\put(0,0){\line(0,1){35}}
\put(15,35){\line(-1,0){15}}
\put(15,35){\line(0,-1){35}}
\put(4,15){\hbox{$C$}}
}
\put(95,27.5){\line(1,0){10}}
\put(95,7.5){\line(1,0){10}}
\put(10,-10){\circle{2}}
\put(10,-10){\circle{1}}
\put(20,-10){\circle{2}}
\put(20,-10){\circle{1}}
\put(30,-10){\circle{2}}
\put(30,-10){\circle{1}}
\put(40,-10){\circle{2}}
\put(40,-10){\circle{1}}
\put(50,-10){\circle{2}}
\put(50,-10){\circle{1}}
\put(60,-10){\circle{2}}
\put(60,-10){\circle{1}}
\put(70,-10){\circle{2}}
\put(70,-10){\circle{1}}
\put(80,-10){\circle{2}}
\put(80,-10){\circle{1}}
\put(90,-10){\circle{2}}
\put(90,-10){\circle{1}}

\end{picture}
\caption{Construction of linear cluster state by repeating the application of $C$ operator.
\label{f:LinClust}}

\end{figure}

Let us start from the linear cluster states. To get the linear cluster state one should first rotate one of qubits from the Bell state $|\Psi^{+}\rangle$ with matrix $\mathcal{B}^{(i)}$:
\begin{equation}
\mathcal{B}^{(i)}=\frac{1}{\sqrt{2}}\left(\begin{array}{cc} 1 & -1 \\ 1 & 1\end{array}\right)
\end{equation}
 Then qubit-2 gets entangled with new qubit-3 via the same operator $\tilde{\mathcal{O}}$ and again should be additionally rotated by the matrix $\mathcal{B}^{(i)}$. Next, qubit-3 entangles with new qubit-4, etc. This gives a linear cluster state with as many qubits as needed. Let us define an operator $C$ for brevity as a composition of entangling operation $\tilde{\mathcal{O}}$ and $B^0$ applied to one of qubits, see Fig.\ref{f:manyb}. Initial state for each qubit is assumed to be the same, $\mathcal{B}^{(i)}|0\rangle$. Then construction of a linear cluster state is made by repeating the operator $C$, like on Fig.\ref{f:LinClust} .

\begin{figure}[h!]
\begin{picture}(200,70)(-175,0)

\put(0,0){
\put(0,0){\line(1,0){15}}
\put(0,0){\line(0,1){15}}
\put(15,15){\line(-1,0){15}}
\put(15,15){\line(0,-1){15}}
\put(2,5){\hbox{$Q_{1}$}}
}
\put(20,0){
\put(0,0){\line(1,0){15}}
\put(0,0){\line(0,1){15}}
\put(15,15){\line(-1,0){15}}
\put(15,15){\line(0,-1){15}}
\put(2,5){\hbox{$Q_{2}$}}
}
\put(40,0){
\put(0,0){\line(1,0){15}}
\put(0,0){\line(0,1){15}}
\put(15,15){\line(-1,0){15}}
\put(15,15){\line(0,-1){15}}
\put(2,5){\hbox{$Q_{3}$}}
}
\put(60,0){
\put(0,0){\line(1,0){15}}
\put(0,0){\line(0,1){15}}
\put(15,15){\line(-1,0){15}}
\put(15,15){\line(0,-1){15}}
\put(2,5){\hbox{$Q_{4}$}}
}
\put(80,0){
\put(0,0){\line(1,0){15}}
\put(0,0){\line(0,1){15}}
\put(15,15){\line(-1,0){15}}
\put(15,15){\line(0,-1){15}}
\put(2,5){\hbox{$Q_{5}$}}
}
\put(100,0){
\put(0,0){\line(1,0){15}}
\put(0,0){\line(0,1){15}}
\put(15,15){\line(-1,0){15}}
\put(15,15){\line(0,-1){15}}
\put(2,5){\hbox{$Q_{6}$}}
}
\put(120,0){
\put(0,0){\line(1,0){15}}
\put(0,0){\line(0,1){15}}
\put(15,15){\line(-1,0){15}}
\put(15,15){\line(0,-1){15}}
\put(2,5){\hbox{$Q_{7}$}}
}
\put(60,40){
\put(0,0){\line(1,0){15}}
\put(0,0){\line(0,1){15}}
\put(15,15){\line(-1,0){15}}
\put(15,15){\line(0,-1){15}}
\put(2,5){\hbox{$Q_{8}$}}
}
{\color{blue}
\put(62,13){
\put(0,0){\line(1,0){10}}
\put(0,0){\line(0,1){29}}
\put(10,29){\line(-1,0){10}}
\put(10,29){\line(0,-1){29}}
\put(1,12){\hbox{$C$}}
}}
\put(15,7.4){\line(1,0){5}}
\put(15,7.6){\line(1,0){5}}
\put(35,7.4){\line(1,0){5}}
\put(35,7.6){\line(1,0){5}}
\put(55,7.4){\line(1,0){5}}
\put(55,7.6){\line(1,0){5}}
\put(75,7.4){\line(1,0){5}}
\put(75,7.6){\line(1,0){5}}
\put(95,7.4){\line(1,0){5}}
\put(95,7.6){\line(1,0){5}}
\put(115,7.4){\line(1,0){5}}
\put(115,7.6){\line(1,0){5}}
\end{picture}
\caption{Example of a simplest two-dimensional (tree) cluster state.
\label{f:TreeC}
}

\end{figure}
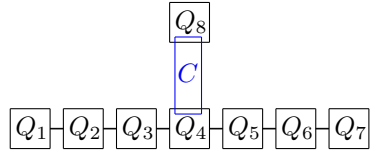

Indeed, it is easy to see, that such state is invariant upon action of correlation operators
\begin{equation}
K^{i}=\sigma_{x}^{i}\bigotimes_{j=i\pm 1}\sigma_{z}^{j}
\end{equation}
This procedure can be easily generalized to get cluster states with non-linear graph. For this operator $C$ should be used to entangle new qubit with one in the middle of a cluster, see Fig.\ref{f:TreeC} for an example. To get this cluster state, first, one creates a cluster state of 7 qubits consecutively acting with operator $C$: $C_{(6)(7)} \cdot C_{(5)(6)}\cdot C_{(4)(5)}\cdot C_{(3)(4)}\cdot C_{(2)(3)}\cdot C_{(1)(2)}\cdot(\mathcal{B}^{(i)}|0\rangle)^{\otimes 7}$. Then one acts with the same operator $C_{(4)(8)}$ on qubits 4 and 8.

\begin{figure}[h!]
\begin{picture}(200,70)(-175,0)

\put(0,0){
\put(0,0){\line(1,0){15}}
\put(0,0){\line(0,1){15}}
\put(15,15){\line(-1,0){15}}
\put(15,15){\line(0,-1){15}}
\put(2,5){\hbox{$Q_{1}$}}
}
\put(20,0){
\put(0,0){\line(1,0){15}}
\put(0,0){\line(0,1){15}}
\put(15,15){\line(-1,0){15}}
\put(15,15){\line(0,-1){15}}
\put(2,5){\hbox{$Q_{2}$}}
}
\put(40,0){
\put(0,0){\line(1,0){15}}
\put(0,0){\line(0,1){15}}
\put(15,15){\line(-1,0){15}}
\put(15,15){\line(0,-1){15}}
\put(2,5){\hbox{$Q_{3}$}}
}
\put(60,0){
\put(0,0){\line(1,0){15}}
\put(0,0){\line(0,1){15}}
\put(15,15){\line(-1,0){15}}
\put(15,15){\line(0,-1){15}}
\put(2,5){\hbox{$Q_{4}$}}
}
\put(80,0){
\put(0,0){\line(1,0){15}}
\put(0,0){\line(0,1){15}}
\put(15,15){\line(-1,0){15}}
\put(15,15){\line(0,-1){15}}
\put(2,5){\hbox{$Q_{5}$}}
}
\put(100,0){
\put(0,0){\line(1,0){15}}
\put(0,0){\line(0,1){15}}
\put(15,15){\line(-1,0){15}}
\put(15,15){\line(0,-1){15}}
\put(2,5){\hbox{$Q_{6}$}}
}
\put(120,0){
\put(0,0){\line(1,0){15}}
\put(0,0){\line(0,1){15}}
\put(15,15){\line(-1,0){15}}
\put(15,15){\line(0,-1){15}}
\put(2,5){\hbox{$Q_{7}$}}
}
\put(60,20){
\put(0,0){\line(1,0){15}}
\put(0,0){\line(0,1){15}}
\put(15,15){\line(-1,0){15}}
\put(15,15){\line(0,-1){15}}
\put(2,5){\hbox{$Q_{8}$}}
}
\put(67.4,15){\line(0,1){5}}
\put(67.6,15){\line(0,1){5}}
\put(67.4,35){\line(0,1){5}}
\put(67.6,35){\line(0,1){5}}

\put(15,7.4){\line(1,0){5}}
\put(15,7.6){\line(1,0){5}}
\put(35,7.4){\line(1,0){5}}
\put(35,7.6){\line(1,0){5}}
\put(55,7.4){\line(1,0){5}}
\put(55,7.6){\line(1,0){5}}
\put(75,7.4){\line(1,0){5}}
\put(75,7.6){\line(1,0){5}}
\put(95,7.4){\line(1,0){5}}
\put(95,7.6){\line(1,0){5}}
\put(115,7.4){\line(1,0){5}}
\put(115,7.6){\line(1,0){5}}

\put(0,40){
\put(0,0){\line(1,0){15}}
\put(0,0){\line(0,1){15}}
\put(15,15){\line(-1,0){15}}
\put(15,15){\line(0,-1){15}}
\put(2,5){\hbox{$Q_{9}$}}
}
\put(20,40){
\put(0,0){\line(1,0){16}}
\put(0,0){\line(0,1){15}}
\put(16,15){\line(-1,0){16}}
\put(16,15){\line(0,-1){15}}
\put(0,5){\hbox{$Q_{10}$}}
}
\put(40,40){
\put(0,0){\line(1,0){16}}
\put(0,0){\line(0,1){15}}
\put(16,15){\line(-1,0){16}}
\put(16,15){\line(0,-1){15}}
\put(0,5){\hbox{$Q_{11}$}}
}
\put(60,40){
\put(0,0){\line(1,0){16}}
\put(0,0){\line(0,1){15}}
\put(16,15){\line(-1,0){16}}
\put(16,15){\line(0,-1){15}}
\put(0,5){\hbox{$Q_{12}$}}
}
\put(80,40){
\put(0,0){\line(1,0){16}}
\put(0,0){\line(0,1){15}}
\put(16,15){\line(-1,0){16}}
\put(16,15){\line(0,-1){15}}
\put(0,5){\hbox{$Q_{13}$}}
}
\put(100,40){
\put(0,0){\line(1,0){16}}
\put(0,0){\line(0,1){15}}
\put(16,15){\line(-1,0){16}}
\put(16,15){\line(0,-1){15}}
\put(0,5){\hbox{$Q_{14}$}}
}
\put(120,40){
\put(0,0){\line(1,0){16}}
\put(0,0){\line(0,1){15}}
\put(16,15){\line(-1,0){16}}
\put(16,15){\line(0,-1){15}}
\put(0,5){\hbox{$Q_{15}$}}
}
\put(15,47.4){\line(1,0){5}}
\put(15,47.6){\line(1,0){5}}
\put(35,47.4){\line(1,0){5}}
\put(35,47.6){\line(1,0){5}}
\put(55,47.4){\line(1,0){5}}
\put(55,47.6){\line(1,0){5}}
\put(75,47.4){\line(1,0){5}}
\put(75,47.6){\line(1,0){5}}
\put(95,47.4){\line(1,0){5}}
\put(95,47.6){\line(1,0){5}}
\put(115,47.4){\line(1,0){5}}
\put(115,47.6){\line(1,0){5}}
\end{picture}
\caption{Example of 15 qubit cluster state, which provides CNOT.
\label{f:TreeCNOT}
}
\end{figure}
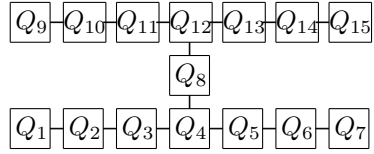

The procedure is flexible and one can also get more complex cluster states, like those, discussed in \cite{Measurement-based}. In fact it is possible to create in such manner a cluster state with any arborescent graph. For example one can construct a 15 qubit cluster state that gives a CNOT gate after measurements, see Fig. \ref{f:TreeCNOT}.

It is obtained as:
\small{
$$C_{(10)(9)} \cdot C_{(11)(10)} \cdot C_{(12)(11)} \cdot C_{(14)(15)} \cdot C_{(13)(14)} \cdot C_{(12)(13)} \cdot C_{(8)(12)} \cdot C_{(4)(8)} \cdot C_{(6)(7)} \cdot C_{(5)(6)}\cdot C_{(4)(5)}\cdot C_{(3)(4)}\cdot C_{(2)(3)}\cdot C_{(1)(2)}\cdot(\mathcal{B}^{(1)}|0\rangle)^{\otimes 15}$$}

And is eigenstate of the corresponding correlation operator
$$K^{i}=\sigma_{x}^{i}\bigotimes_{j\in nbgh(i)}\sigma_{z}^{j}$$

The generalization of this approach to clusters with non-tree graphs is non-trivial and remains to be studied.

\section{Conclusion}

In this paper we discussed how arborescent knots can be used to construct multi-qubit states. In particular we've shown that using particular four-strand braids one can get Bell-states -- maximally entangled two-qubit state. By further applications of same approach GHZ-states and even cluster states can be constructed. Complex cluster states can be used to construct a measurement-based quantum computer.

It is interesting to further understand which cluster states can be constructed using such approach, for example, if it is possible to get cluster state with non-tree graph. Another question is how to increase probability of applying these two-qubit operation. At the moment the described approach works by applying the procedure until all the required anyons annihilate, but this seriously hampers the probability of applying this operation or in other words its fidelity. Thus it is interesting how to modify this operation to get higher fidelity.

\section*{Aknowledgements}

The authors are grateful for very useful discussions with I.Diakonov and N.Kolganov. This work was supported by the Russian Science Foundation grant No 23-71-10058.




\begin{thebibliography}{99}

\bibitem{Kit}
A.~Y.~Kitaev,
Russian Mathematical Surveys, 52 (1997), 1191.

\bibitem{ReviewTQC}
C.~Nayak, S.~H.~Simon, A.~Stern, M.~Freedman, S.~Das Sarma,
Rev. Mod. Phys. 80 (2008), 1083,
[arXiv:0707.1889 [hep-th]].

\bibitem{exp1} R.~S.~Souto, A.~Tsintzis, M.~Leijnse, J.~Danon,
Phys. Rev. Research 5, 043182 (2023)
doi:10.1103/PhysRevResearch.5.043182
[arXiv:2308.14751 [cond-mat]].

\bibitem{exp2} D.~Zou, N.~Pan, T.~Chen, H.~Sun, X.~Zhang,
Adv. Intell. Syst. 2300354 (2023)
[arXiv:2309.04896 [cond-mat]].

\bibitem{exp3} A.~Winblad, H.~Chen,
[arXiv:2309.11607 [cond-mat]].

\bibitem{exp4} P.~Bonderson, A.~Kitaev, K.~Shtengel,
 Phys. Rev. Lett. 96, 016803 (2006)
[arXiv:cond-mat/0508616].

\bibitem{exp5} J.~Nakamura, S.~Liang, G.~C.~Gardner, M.~J.~Manfra,
Nat. Phys. 16, 931–936 (2020),
doi:10.1038/s41567-020-1019-1
[arXiv:2006.14115 [cond-mat]].

\bibitem{expend} F.~Domínguez, F.~Hassler, G.~Platero,
[arXiv:1202.0642 [cond-mat]].

\bibitem{KLJ}  L.~Kauffman, S.~Lomonaco, New Journal of Physics, \textbf{4} (2002) 73.1-18; 6 (2004) 134.1-40, [arXiv:quant-ph/0401090].

\bibitem{TopoBook} S.~H.~Simon, Topological Quantum (Oxford, 2023; online edn, Oxford Academic, 14 Dec. 2023).

\bibitem{TowTQC}
D.~Melnikov, A.~Mironov, S.~Mironov, A.~Morozov and An.~Morozov,
Nucl. Phys. B \textbf{926} (2018), 491-508
doi:10.1016/j.nuclphysb.2017.11.016
[arXiv:1703.00431 [hep-th]].

\bibitem{QuantKnots}
N.~Kolganov and An.~Morozov,
JETP Lett. \textbf{111} (2020) no.9, 519-524
doi:10.1134/S0021364020090027
[arXiv:2004.07764 [hep-th]].

\bibitem{LargeK}
N.~Kolganov, S.~Mironov and An.~Morozov,
Nucl. Phys. B \textbf{987} (2023), 116072
doi:10.1016/j.nuclphysb.2023.116072
[arXiv:2105.03980 [hep-th]].

\bibitem{Measure}
An.~Morozov, Problems of Information Transmission 60, 28-34 (2024)
doi:10.1134/S0032946024010046
[arXiv:2403.07847 [hep-th]].

\bibitem{Bel}
A.~Belov, An.~Morozov,
[arXiv:2408.14188 [hep-th]].

\bibitem{clusterstate}
R.~Raussendorf and H.~Briegel,
Phys. Rev. Lett. \textbf{86}, 910 (2001)
doi:10.1103/PhysRevLett.86.910
[arXiv:quant-ph/0004051].

\bibitem{QCc}
R.~Raussendorf and H.~Briegel,
Phys. Rev. Lett. \textbf{86}, 5188 (2001)
doi:10.1103/PhysRevLett.86.5188
[arXiv:quant-ph/0108067].

\bibitem{Measurement-based}
R.~Raussendorf, D.~Browne and H.~Briegel,
Phys. Rev. A. \textbf{68}, 022312 (2003)
doi:10.1103/PhysRevA.68.022312
[arXiv:quant-ph/0301052].

\bibitem{Tabul}
A.~Mironov, A.~Morozov, An.~Morozov, P.~Ramadevi, V.~K.~Singh and A.~Sleptsov,
J. Phys. A \textbf{50} (2017) no.8, 085201
doi:10.1088/1751-8121/aa5574
[arXiv:1601.04199 [hep-th]].

\bibitem{DoubleFat}
A.~Mironov, A.~Morozov, An.~Morozov, P.~Ramadevi and V.~K.~Singh,
JHEP \textbf{07} (2015), 109
doi:10.1007/JHEP07(2015)109
[arXiv:1504.00371 [hep-th]].

\bibitem{NPZ}  S.~Nawata, P.~Ramadevi, Zodinmawia,
Letters in Mathematical Physics \textbf{103}, pp. 1389–1398 (2013), [arXiv:1302.5143 [hep-th]]

\bibitem{MMS} A. Mironov, A. Morozov, A. Sleptsov,
JHEP \textbf{07} (2015) 069, [arXiv:1412.8432 [hep-th]]


\bibitem{UniGate}
C.~Levaillant, B.~Bauer, M.~Freedman, Z.~Wang, P.~Bonderson,
 	Phys. Rev. A \textbf{92}, 012301 (2015)
doi:10.1103/PhysRevA.92.012301
[arXiv:1504.02098 [quant-ph]].


\bibitem{HMFLII}
A.~Mironov, A.~Morozov and An.~Morozov,
JHEP \textbf{03} (2012), 034
doi:10.1007/JHEP03(2012)034
[arXiv:1112.2654 [math.QA]].

\bibitem{Cable}
A.~Anokhina and An.~Morozov,
Teor. Mat. Fiz. \textbf{178} (2014), 3-68
doi:10.1007/s11232-014-0129-2
[arXiv:1307.2216 [hep-th]].

\bibitem{RPaths}
A.~Anokhina, A.~Mironov, A.~Morozov and An.~Morozov,
Adv. High Energy Phys. \textbf{2013} (2013), 931830
doi:10.1155/2013/931830
[arXiv:1304.1486 [hep-th]].

\bibitem{MultiLink}
S.~Dhara, A.~Mironov, A.~Morozov, An.~Morozov, P.~Ramadevi, V.~K.~Singh and A.~Sleptsov,
Annales Henri Poincare \textbf{20} (2019) no.12, 4033-4054
doi:10.1007/s00023-019-00841-z
[arXiv:1805.03916 [hep-th]].

\bibitem{MultiLink1}
C.~Bai, J.~Jiang, J.~Liang, A.~Mironov, A.~Morozov, An.~Morozov and A.~Sleptsov,
J. Geom. Phys. \textbf{132} (2018), 155-180
doi:10.1016/j.geomphys.2018.05.020
[arXiv:1801.09363 [hep-th]].

\end{thebibliography}
\end{document}